\documentclass[a4paper]{jpconf}
\usepackage[latin1]{inputenc}
\usepackage{graphicx}
\usepackage{subfigure}
\usepackage{xcolor}
\usepackage{amsmath}
\usepackage[initials]{amsrefs}
\usepackage{amssymb}
\usepackage{bbold}
\usepackage{slashed}
\usepackage{tabularx}

\allowdisplaybreaks[1]

\newcommand{\MeV}{\,\mathrm{MeV}}

\newcommand{\Lag}{\mathcal{L}}
\newcommand{\LSM}{\mathrm{L}\sigma\mathrm{M}}
\newcommand{\LLSM}{\mathcal{L}_{\mathrm{L}\sigma\mathrm{M}}}

\newcommand{\ie}{\textit{i.e.\,}}
\newcommand{\eg}{\textit{e.g.\,}}
\newcommand{\cf}{\textit{cf.\,}}
\newcommand{\closeparentheses}[1]{#1)}
\newcommand{\Ord}[1]{\mathcal{O}(#1)}

\allowdisplaybreaks[1]

\BibSpec{article}{+{}{\PrintAuthors}{author}+{}{\ }{year}+{, }{\it}{journal}+{\ }{\textbf}{volume}+{, }{}{pages}}
\BibSpec{book}{+{}{\PrintAuthors}{editor}+{}{\ (eds.)\ \it}{title}+{}{,\ }{series}+{}{\ \textbf}{volume}+{ }{\ (}{publisher}+{}{,\ }{address}+{ }{,\ \closeparentheses}{year}}
\begin{document}
\title{Photon emission within the linear sigma model}
\author{F Wunderlich and B K\"ampfer}
\address{Helmholtz-Zentrum Dresden-Rossendorf 01328 Dresden, Germany and}
\address{Institut f\"ur Theoretische Physik, TU Dresden 01062 Dresden, Germany}
\ead{f.wunderlich@hzdr.de, b.kaempfer@hzdr.de}

\begin{abstract}
   Soft-photon emission rates are calculated within the linear sigma model.
   The investigation is aimed at answering the question to which extent the emissivities map out the phase structure
   of this particular effective model of strongly interacting matter.
\end{abstract}

\section{Introduction}
Despite of the increasing success of QCD in describing a large variety of phenomena, both in the perturbative
as well as in the non-perturbative regimes, some fundamental questions remain unsolved. Prominent examples are the 
very nature and detailed properties of the strongly coupled quark gluon plasma which is the conjectured state of QCD
matter at temperatures comparable and larger than the QCD energy scale $\Lambda_{\mathrm{QCD}}$. 
Furthermore, the nature and properties of the chiral and deconfinement phase transition as well as the position 
of a conjectured critical point (CP) in the QCD phase diagram are among the still challenging issues \cite{Friman:2011zz}.
To answer these questions experimentally,  a number of large scale experiments is currently running 
(ALICE, ATLAS and CMS at the LHC, and STAR and PHENIX at RHIC),
planned (MPD at NICA) or under construction (CBM and HADES at FAIR).

On the theory side, lattice QCD yields a smooth
crossover from the hadronic phase to the quark-gluon phase for small chemical potential at temperatures of about 
$150\MeV$. At sufficiently large net baryon density the crossover may turn into a first order phase transition 
at a CP.
At non-zero net densities (\ie non-zero quark chemical potential) there is no first-principle approach to the phase diagram. 
Therefore, one has to rely on effective models or on 
truncation schemes. Nevertheless, many of these approaches seem to point to a first order phase transition
connected to the spontaneous breaking of chiral symmetry at densities a few times the nuclear density. 

The end point of this transition line has interesting properties on its own. 
From macroscopic examples, the phenomenon of critical opalescence, \ie the diverging
scattering strength of transparent media in the vicinity of a critical point, has been known for a long time 
\cite{Smoluchowski:1908}. Quite common also is the phenomenon of critical slowing down, \ie the diverging relaxation 
time at criticality \cite{Hohenberg:1977ym}. These two examples as well as most of the special properties of critical points
have their reason in the diverging correlation length making the system scale free.

To understand the mass generation connected to chiral symmetry breaking, several effective models have been constructed
with the Nambu-Jona-Lasinio (NJL) model \cites{Klevansky:1992qe,Asakawa:1989bq} and the 
linear sigma model (L$\sigma$M) \cites{Bochkarev:1995gi, Jungnickel:1995fp} being the most prominent ones.

Our present investigation is motivated by the question whether penetrating probes reflect directly the phase structure
of strongly interacting matter. We  focus here on real photons and select the $\LSM$ to mimic the above 
anticipated phase structure. The $\LSM$ contains quark and meson (pion and sigma) fields as basic degrees of freedom, where the fluctuations
of the latter ones are accounted for in linear approximation, as in \cites{Mocsy:2004ab,Bowman:2008kc,Ferroni:2010ct} and 
the photon field is minimally coupled to the strongly interacting components of the $\LSM$.

There is a large difference in the time scales concerning the strong and the electromagnetic interactions, respectively.
This makes possible separating the two interactions involved. The strong interaction is responsible for the relaxation
towards a local thermal equilibrium as well as to the mass generation via the spontaneous breaking of chiral symmetry.
The electromagnetic interaction with a perturbative radiation field contributes little to this, 
because its effects are $ \Ord{\alpha_{\mathrm{em}}/\alpha_s }$ 
suppressed. Therefore we might calculate the thermodynamics without regarding electromagnetism and use the thermodynamic
properties as well as the effective masses of the dressed quarks and mesons later on in the photon emission calculations.

\section{Thermodynamics of the L$\sigma$M with linearized meson fluctuations}
The $\LSM$ is a widely used effective model of QCD and has been applied often
for studying various aspects of thermodynamics of strongly interacting matter.
It was suggested by Gell-Mann and Levy in 1960 \cite{GellMann:1960np} for 
studying chiral symmetry breaking. In absence of an explicit symmetry breaking term, the model has a 
\mbox{$SU(2)\times SU(2)\simeq O(4)$} symmetry and therefore belonging to the same 
universality class as $N_f=2$ QCD in the chiral limit \cite{Pisarski:1983ms}. This symmetry present
at high temperatures is spontaneously broken to a residual $SU(2)$ symmetry with the three pseudoscalar $\pi$ mesons 
being the Goldstone modes. Breaking chiral symmetry explicitly, the pions acquire non-zero masses. Besides these
satisfying properties
there is a close connection to the non-linear $\sigma$ model, which in turn is equivalent to leading order 
chiral effective field theory of QCD. Compared to the NJL model the $\LSM$ has the advantage of including the mesons
directly as dynamic field quanta, making it easier to address their properties.

The $\LSM$ Lagrangian reads
\begin{eqnarray}
   \LLSM &=& \bar\psi(i\gamma^\nu\partial_\nu - g(\sigma + i\gamma^5\vec\tau\vec \pi))\psi 
                + \frac12\partial_\rho \sigma \partial^\rho \sigma + \frac12\partial_\kappa\vec\pi\partial^\kappa \vec\pi
                + \frac{\lambda}{4}(\sigma^2 + \vec\pi^2 - v^2)^2 - H \sigma,
\end{eqnarray}
where the Dirac field $\psi$ describes a doublet of quarks,
$\sigma$ corresponds to an iso-scalar and Lorentz-scalar field,
and $\vec{\pi}$ describes an iso-vector and Lorentz-pseudoscalar field, the latter ones conveniently interpreted as the 
$\sigma$ and $\pi$ mesons.
From the Lagrangian the thermodynamic 
potential $\Omega$ is constructed via the path integral of the exponential of the Euclidean action
and evaluated following the procedure described 
in \cites{Mocsy:2004ab,Bowman:2008kc,Ferroni:2010ct} for including linearized fluctuations.
First, one integrates over the fermionic fields $\psi$ and $\bar\psi$. 
The remaining path integral corresponds to a purely mesonic theory with a complicated interaction potential, which
is approximated by a quadratic one to account for small fluctuations. 
The parameters of this quadratic potential are identified with the masses and thermodynamic averages of the
meson fields. This leads to self consistency relations for the masses. 

The parameters are fixed by the following requirements: 
The mass of the pions is set to $138\MeV$ in vacuum
($T=\mu=0$) and the sigma meson mass to $700\MeV$. The effective quark mass in the vacuum 
is fixed to one third of the nucleon mass $m_{\mathrm{eff}}^0 = g v = 312\MeV$, and the parameter $v$ is 
identified with the pion decay constant in vacuum, $v=92.4\MeV$.

\begin{figure}[htp]
   \centering
   \subfigure{\includegraphics[width = 0.47\textwidth,clip=true,trim=7mm 12mm 30mm 25mm]
             {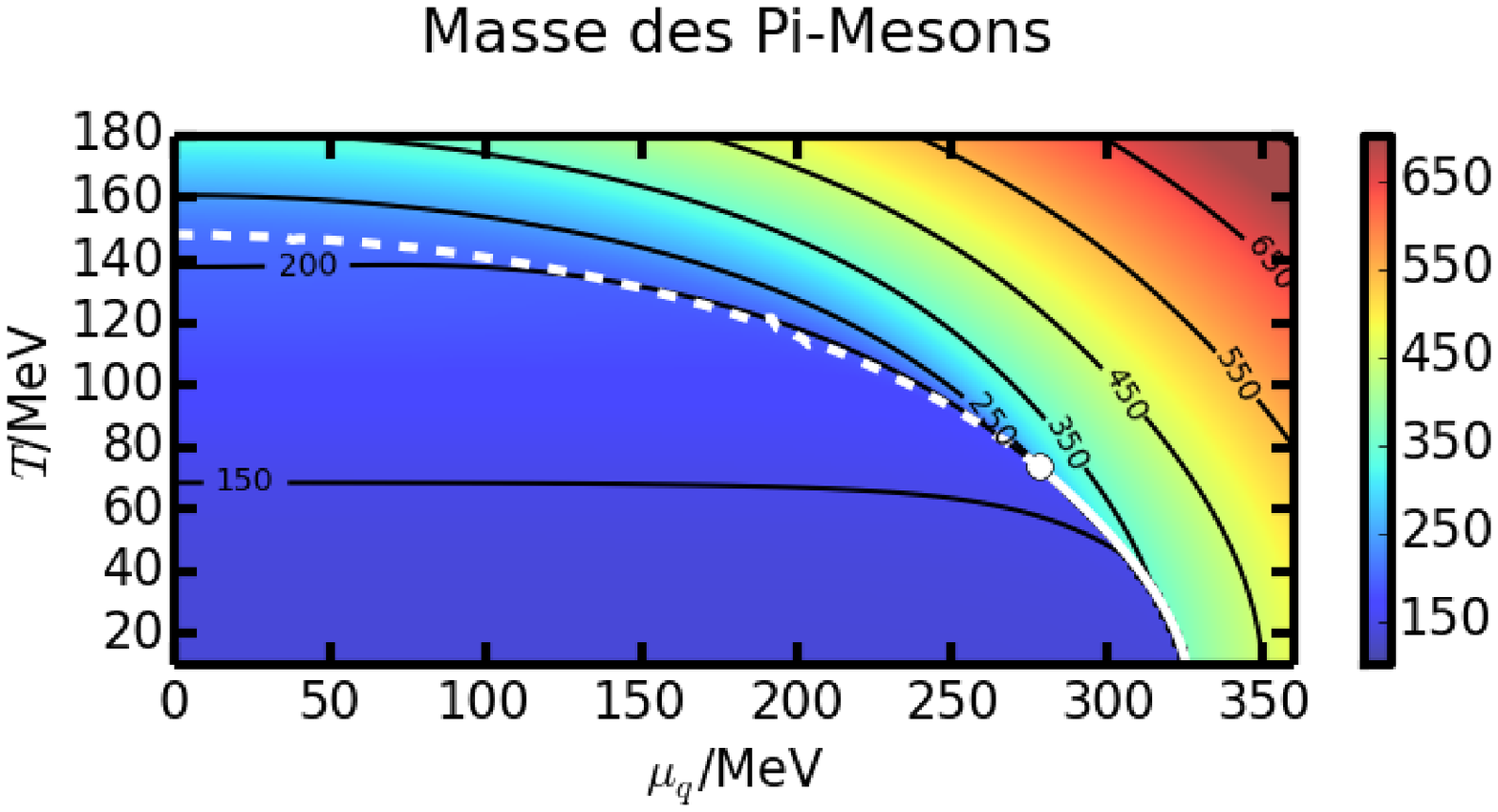}
             \label{subfig_m_pi_LF}
             \put(-191, 30){\fcolorbox{black}{white}{(a)}}
             }
   \subfigure{\includegraphics[width = 0.47\textwidth,clip=true,trim=7mm 12mm 30mm 25mm]
             {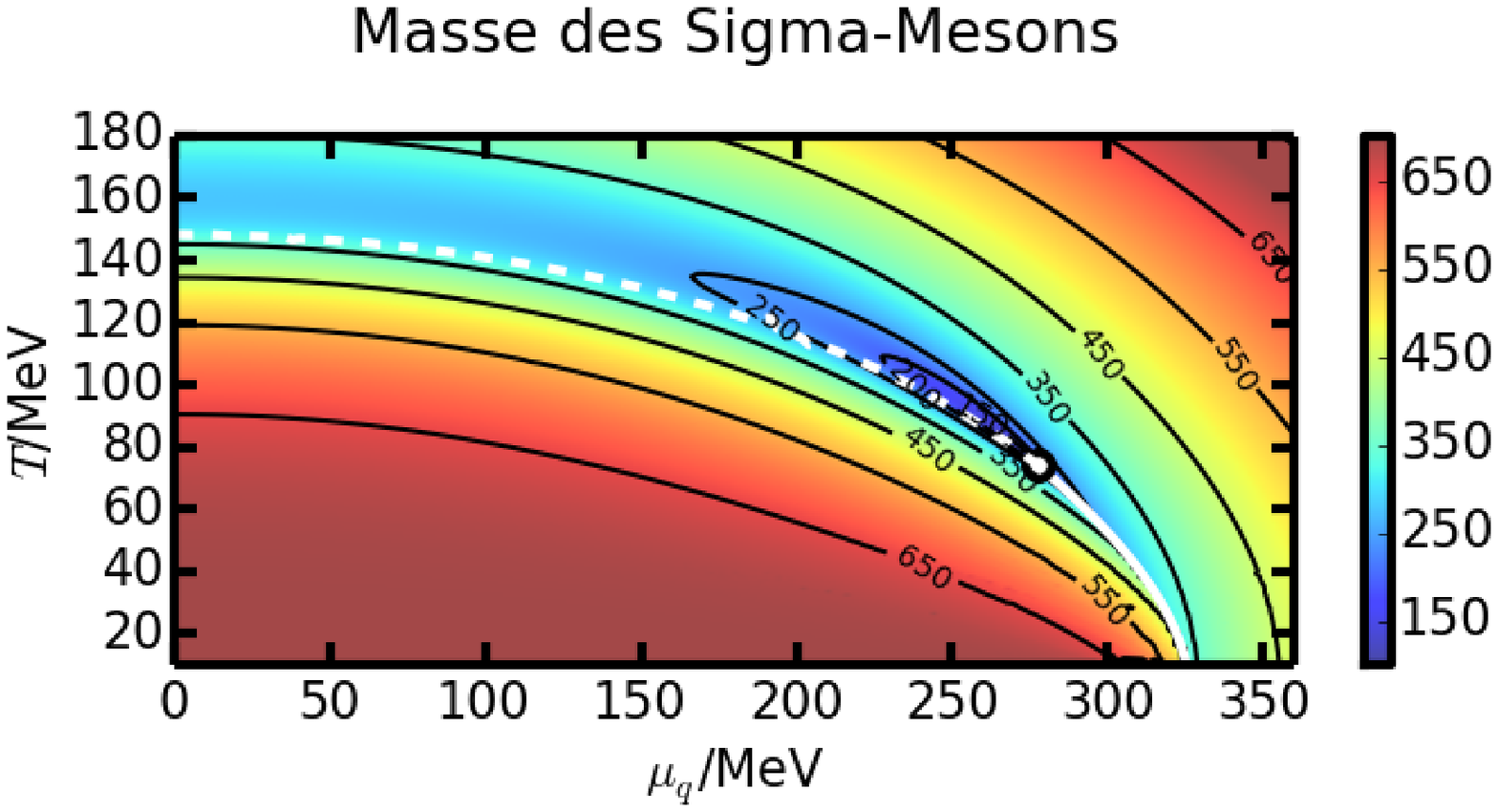}
             \label{subfig_m_si_LF}
             \put(-191, 30){\fcolorbox{black}{white}{(b)}}}\\
   \subfigure{\includegraphics[width = 0.47\textwidth,clip=true,trim=7mm 12mm 30mm 25mm]
             {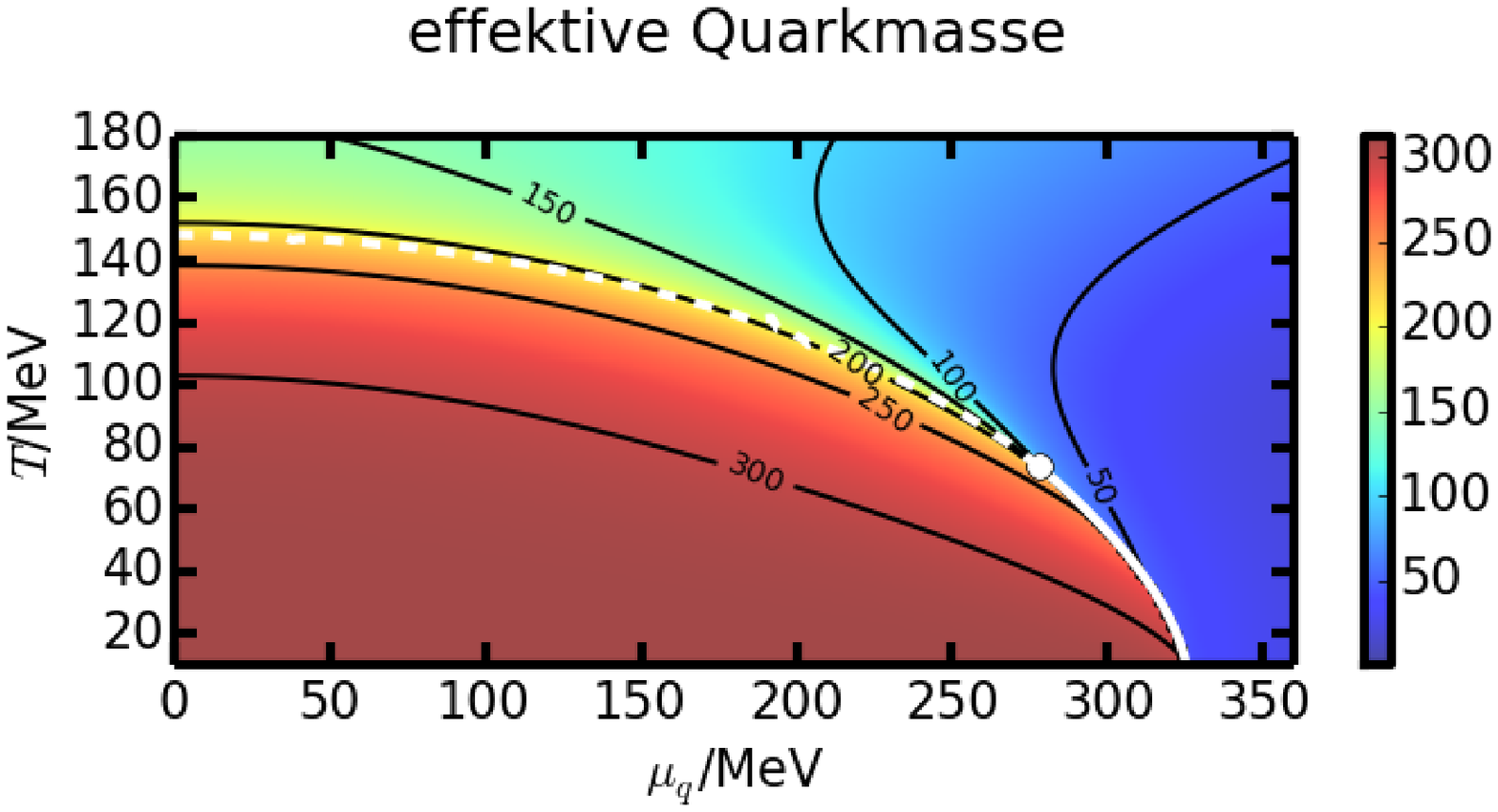}
             \label{subfig_m_q_LF}
             \put(-191, 30){\fcolorbox{black}{white}{(c)}}
             }
   \subfigure{\includegraphics[width = 0.47\textwidth,clip=true,trim=7mm 12mm 30mm 25mm]
             {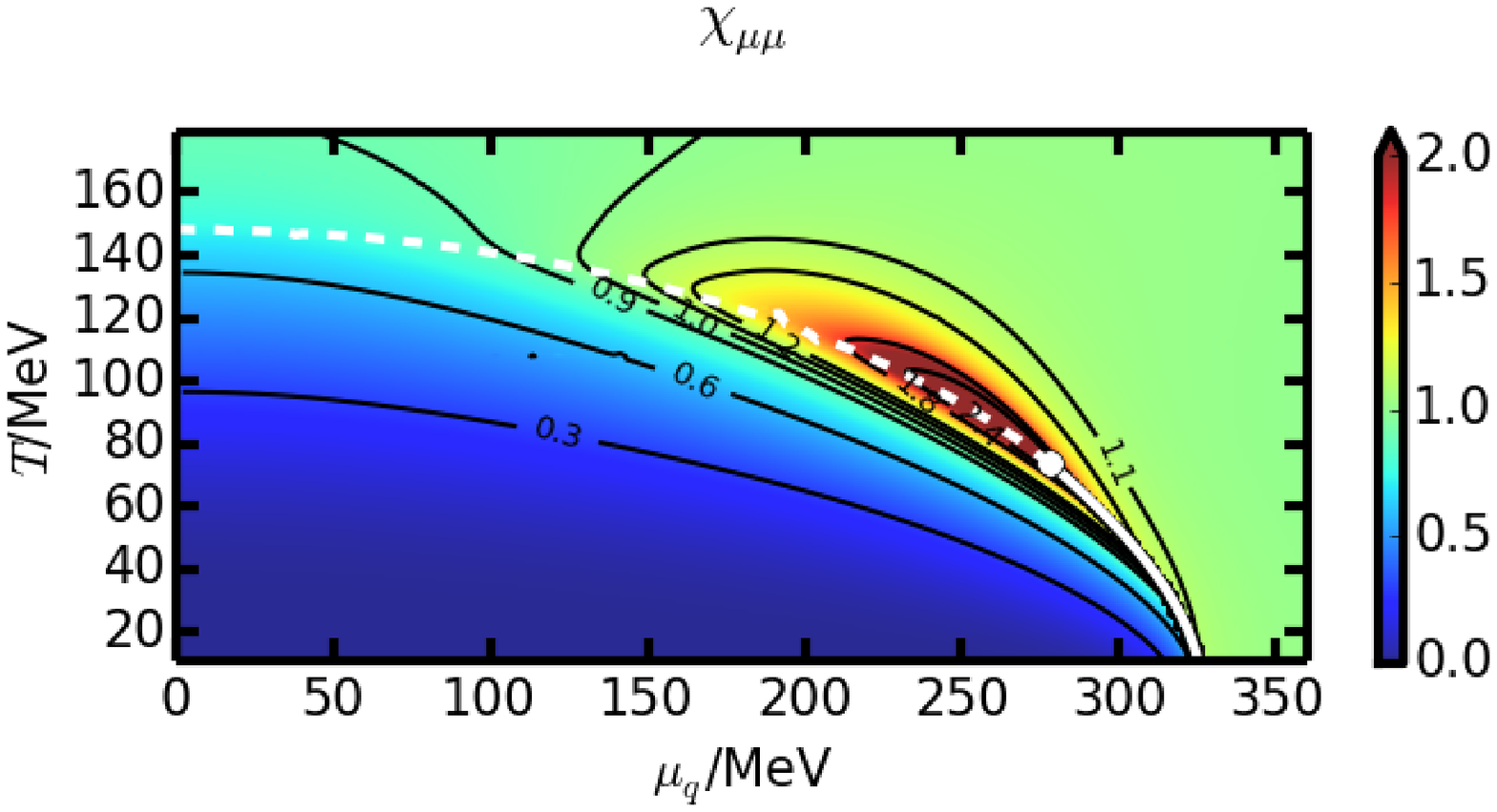}
             \label{subfig_chi_LF}
             \put(-191, 30){\fcolorbox{black}{white}{(d)}}
             }
   \caption{Contour plots of dynamically generated masses for the pion (a) and sigma (b) mesons as well as
            the quarks (c) (in MeV).
            The quark number susceptibility (d), normalized to the susceptibility of a ideal massless Fermi-gas,
            is increased around the CP (white circle).
            The solid white curve denotes the coexistence curve for the 1st order phase transition
            and the dashed line estimates the pseudocritical temperature. The later one is defined as the 
            the position of the extremal normalized heat capacity $\bar c = -c_0^{-1}\partial^2_T \Omega(\mu,T)$ 
            with $-c_0$ being the second derivative w.r.t. temperature of the grand canonical potential $\Omega_0$ 
            of an ultrarelativistic free Fermi gas.
            }
   \label{fig_Thermodyn}
\end{figure}

With these parameters one obtains the results depicted in Fig.~\ref{fig_Thermodyn}.
Figs.~\ref{subfig_m_pi_LF}-\ref{subfig_m_q_LF} show contour plots of the masses over the phase diagram. 
One notes that the pion mass (Fig.~\ref{subfig_m_pi_LF}) increases with temperature and chemical potential 
with the strongest
change at the phase boundary. The sigma meson mass (Fig.~\ref{subfig_m_si_LF}) on
the other hand exhibits a valley of low mass values around
the phase boundary and with a global minimum at the critical point. The quark mass plotted in Fig.~\ref{subfig_m_q_LF}
drops from its vacuum value to about $30\MeV$. The most drastic change, again, is at the phase boundary, 
signaling that the mechanism for mass generation is indeed the spontaneous breaking
of chiral symmetry within the $\LSM$. Because the chiral symmetry is also explicitly
broken by a nonzero $H$ in the Lagrangian, the quark mass does not drop to zero, but stays finite in the high 
temperature phase.
Comparing the meson masses (\cf Figs.~\ref{subfig_m_pi_LF} and \ref{subfig_m_si_LF}), one realizes that they are
degenerate above the 1st order phase transition curve and the crossover region, 
respectively, but very different below. This behavior of the mass difference of these chiral partners is another
sign of the chiral symmetry breaking and restoration.

For quantifying the size of the critical region the quark number susceptibility 
\mbox{$\chi:=-\partial^2\Omega/\partial \mu^2$}
is chosen, since the susceptibility scales with the correlation length whose divergence causes many of the special features
of a CP. In Fig.~\ref{subfig_chi_LF}, $\chi$ is normalized to the susceptibility $\chi_0$
of a massless ideal fermion gas to scale out trivial contributions. 

\section{Photon emission rates within the L$\sigma$M}
For calculating photon emission rates, the $\LSM$ Lagrangian is extended by an electromagnetic 
sector coupled minimally (\cf \cite{Mizher:2010zb}) to the strongly interacting part.
\begin{eqnarray}
   \Lag_{\gamma\mathrm{L}\sigma\mathrm{M}} 
         &=& \LLSM + \Lag_\gamma + \Lag_{\mathrm{int}},\\
   \Lag_{\mathrm{int}} &=& -eQ_f\bar\psi \slashed A \psi 
                           + \frac12 e^2 \pi^+\pi^-A^\nu A_\nu 
                           + \frac12 e A_\nu(\pi^-\partial^\nu\pi^+ + \pi^+\partial^\nu\pi^-),
\end{eqnarray}
where $\Lag_\gamma = -\frac14 F^{\mu\nu}F_{\mu\nu}$ is the free photon Lagrangian and $A^\mu$ denotes the photon field.

Photon emission rates are, in a kinetic theory approach, convolutions of squared matrix elements $|M|^2$ and 
phase space distribution functions $f_\pm$, the latter ones explicitly depending on $T$ and $\mu$. 
Superimposed are implicit $T$ and $\mu$ dependencies from the effective masses of the involved fields, as displayed in 
\mbox{Figs.~\ref{subfig_m_pi_LF}-\ref{subfig_m_q_LF}}. Given the marked variations of these masses
one can expect an pronounced impact on the emission rates
\begin{eqnarray}
   \omega \frac{d^7 N}{dx^4dk^3} = \frac{\mathcal{N}}{(2\pi)^5} \int \frac{dp^3}{2p^0}\int \frac{dq^3}{2q^0}\int \frac{dz^3}{2z^0}
                           |\mathcal{M}|^2 f_\pm(p^0)f_\pm(q^0)(1\mp f_\pm(z^0))\delta^{(4)}(p+q-z-k).\label{rate_allg}
\end{eqnarray}

Owing to the weakness of the electromagnetic interaction we restrict the calculations to
first order in the electromagnetic coupling. Since we expect to have captured the dominant part of the strong interaction
in the calculation of the thermodynamic potential and the effective masses, the residual interaction
is expected to be relatively weak. 
Therefore we restrict our calculation to 1st order processes in the quark-meson coupling. 
Within this approximation the contributing processes are the tree-level processes in the $s$, $t$ and $u$ channels.

\begin{figure}[hbt]
   \centering
   \subfigure{\includegraphics[width=0.47\textwidth,clip=true,trim=7mm 12mm 23mm 25mm]{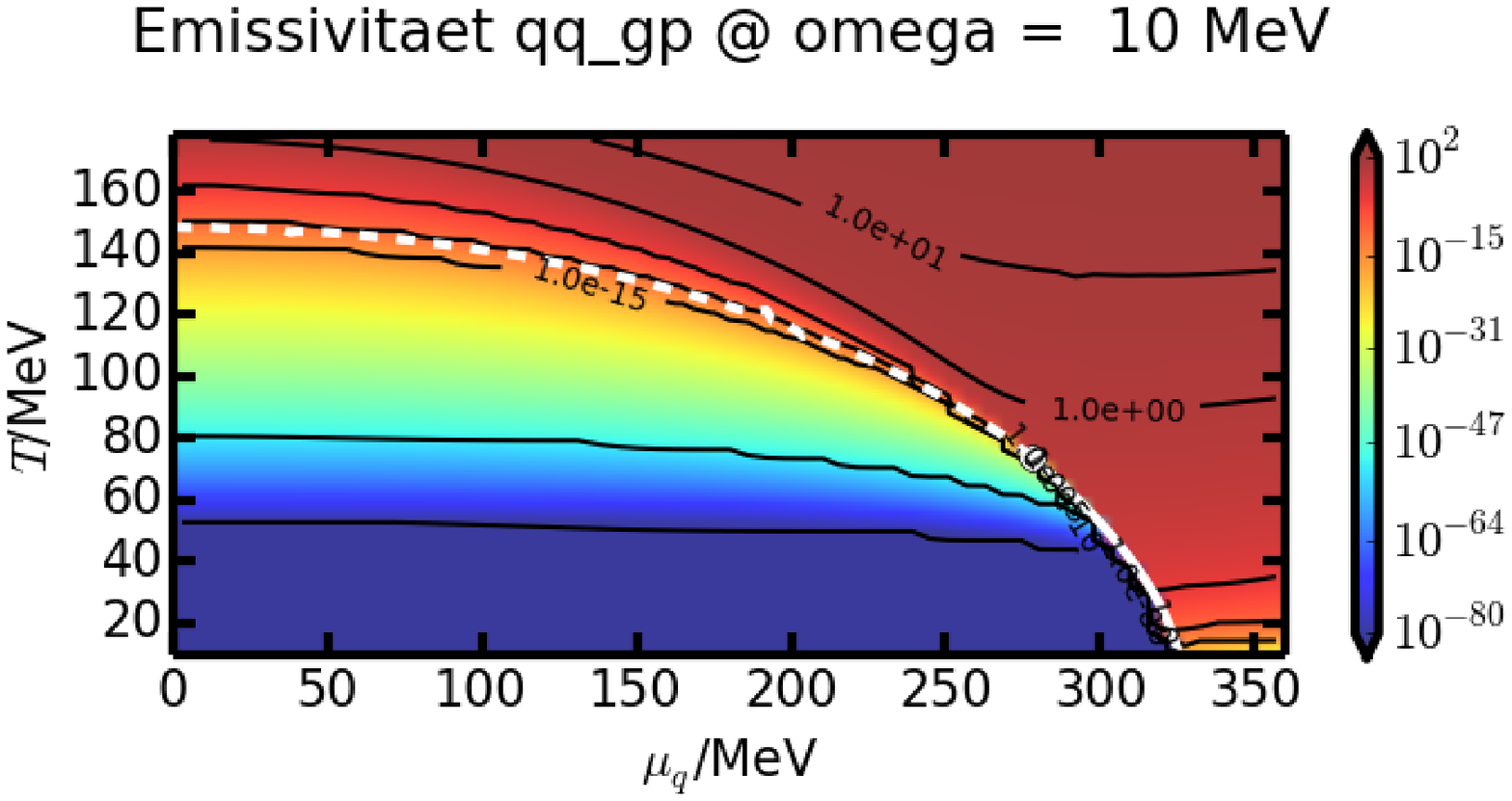}
             \put(-60, 77){\fcolorbox{black}{white}{(a)}}
             \label{subfig_rate_qq_gp_omega=0010}}\hfill
   \subfigure{\includegraphics[width=0.47\textwidth,clip=true,trim=7mm 12mm 23mm 25mm]{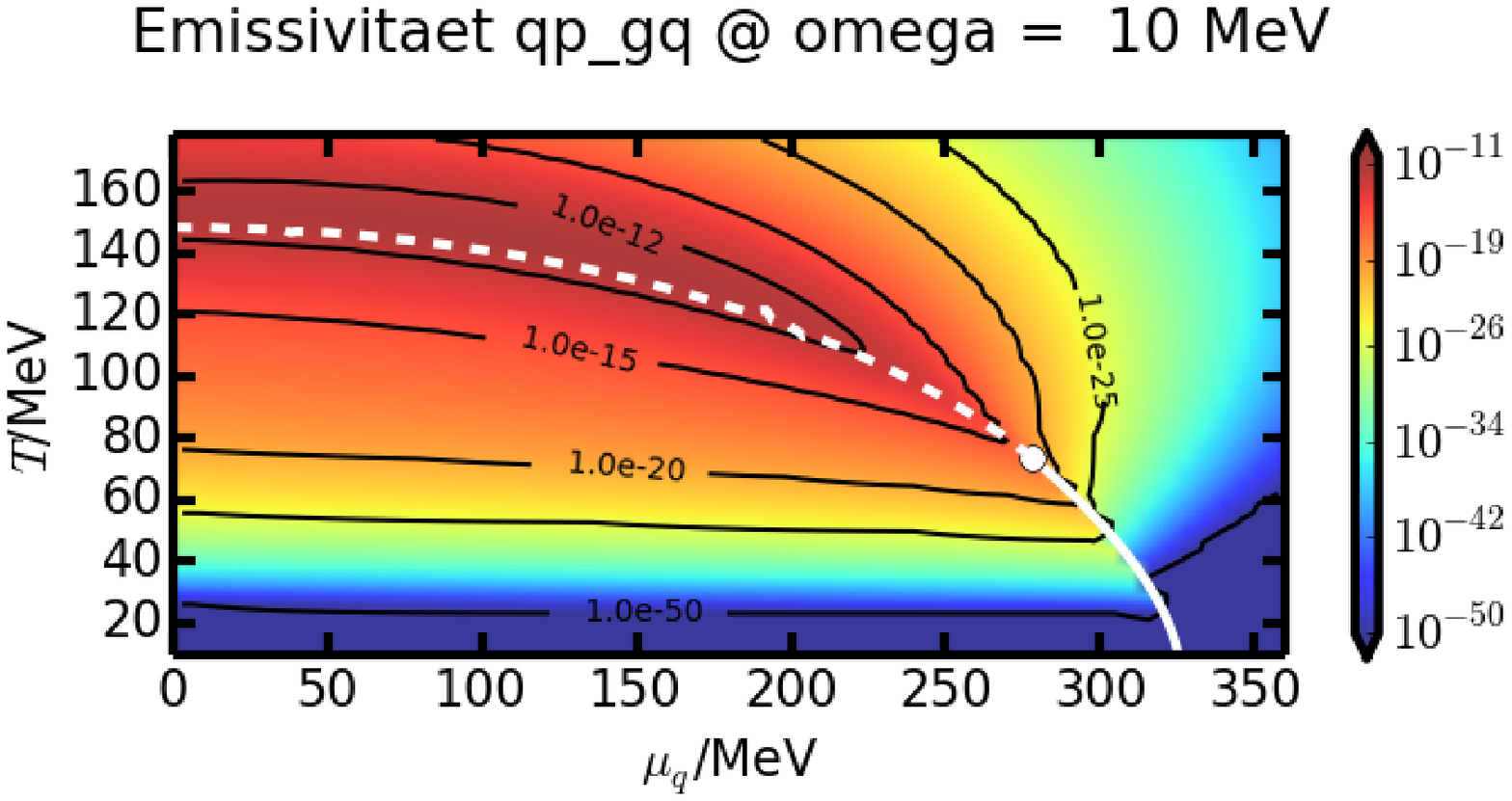}
             \put(-60, 77){\fcolorbox{black}{white}{(b)}}
             \label{subfig_rate_qp_gq_omega=0010}}\\
   \subfigure{\includegraphics[width=0.47\textwidth,clip=true,trim=7mm 12mm 23mm 25mm]{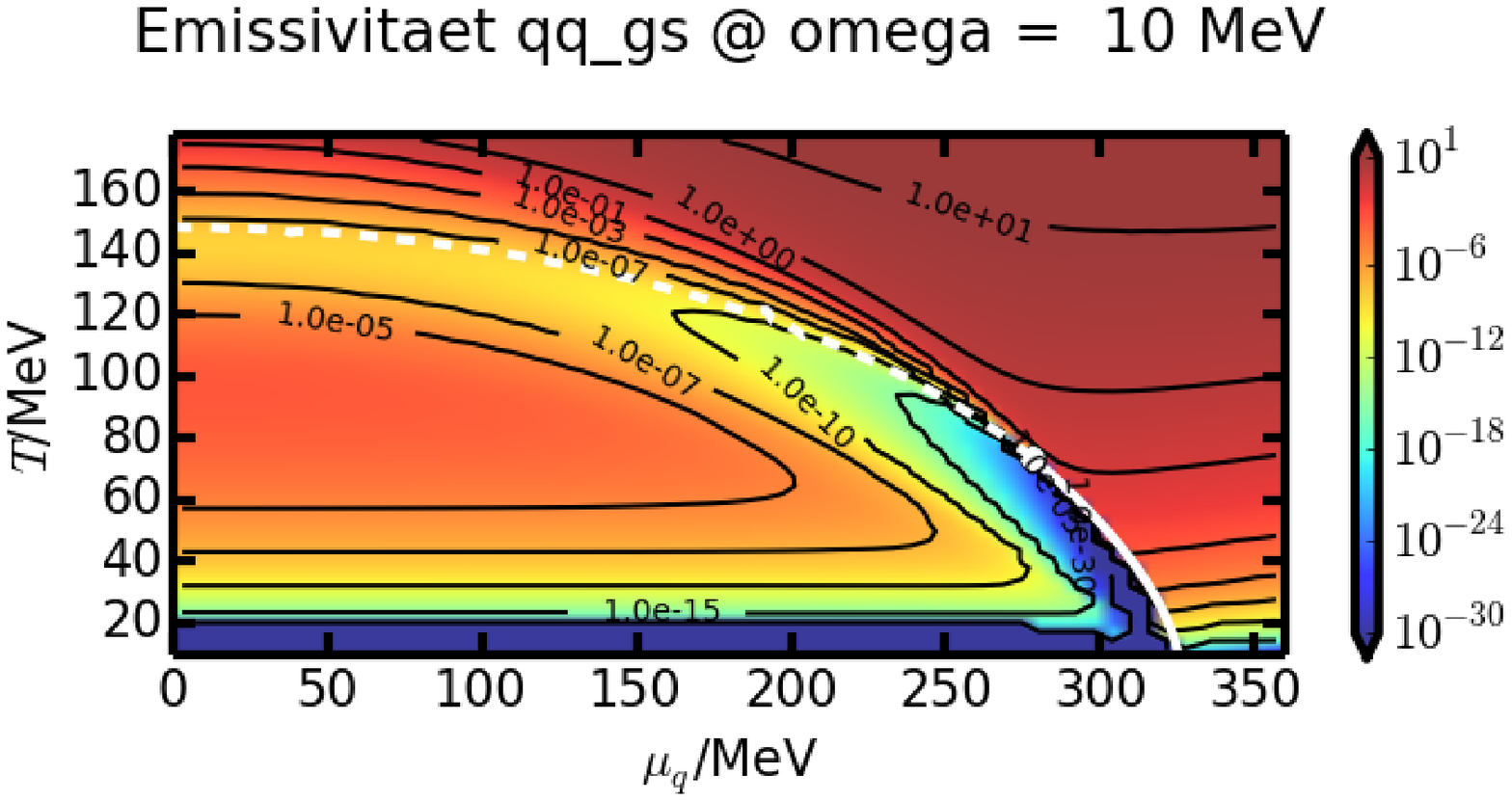}
             \put(-60, 77){\fcolorbox{black}{white}{(c)}}
             \label{subfig_rate_qq_gs_omega=0010}}\hfill
   \subfigure{\includegraphics[width=0.47\textwidth,clip=true,trim=7mm 12mm 23mm 25mm]{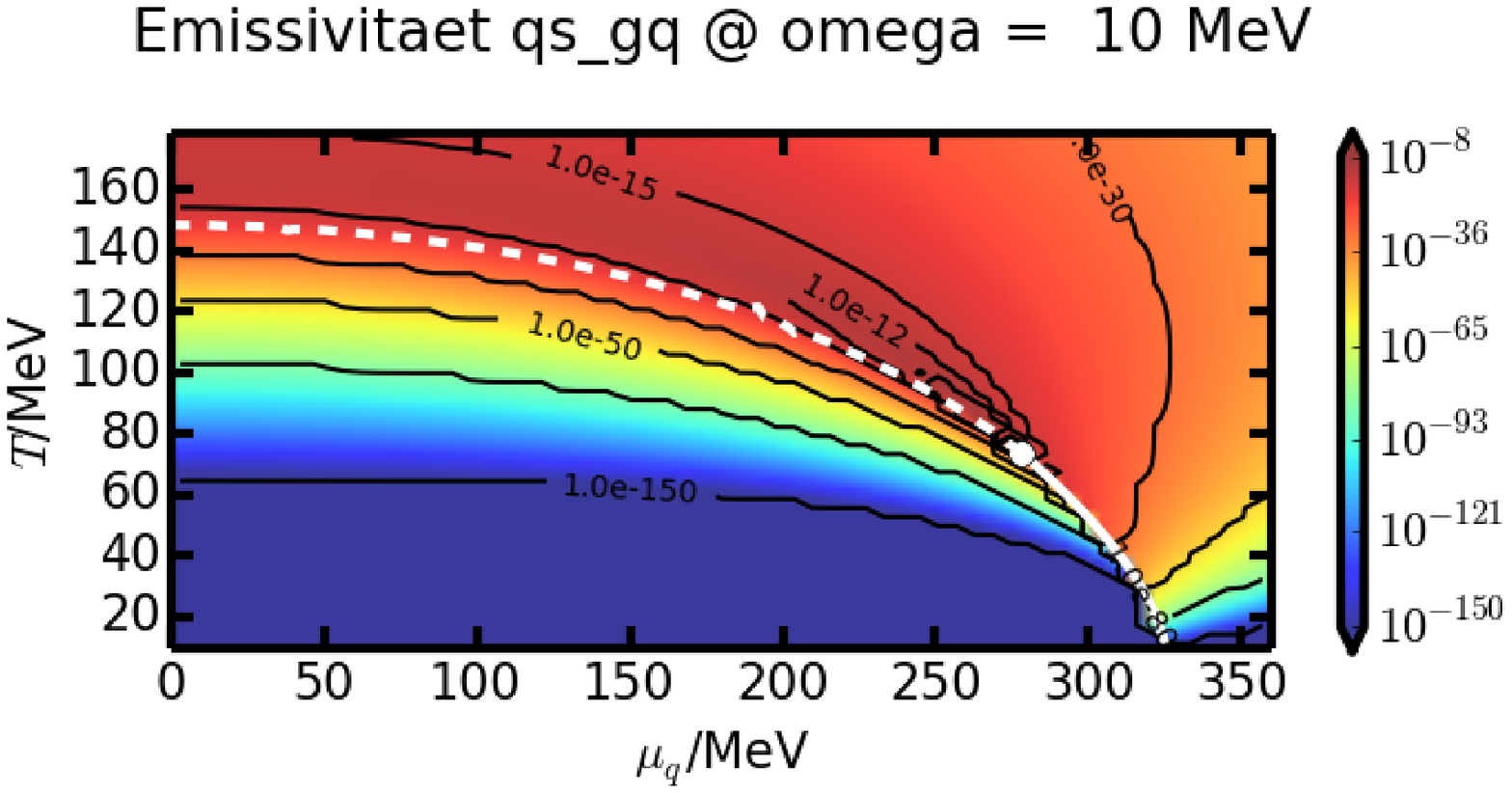}
             \put(-60, 77){\fcolorbox{black}{white}{(d)}}
             \label{subfig_rate_qs_gq_omega=0010}}
    \caption{Contour plots of the photon emissivity $\omega\frac{d^7N}{d^4x d^3k}$ in units of $\MeV^2$ as functions of 
             $T$ and $\mu$ at photon energies $\omega = 10\MeV$ for the processes 
             \mbox{$\overline{\psi} \psi\rightarrow \pi\gamma$ (a)}, 
             \mbox{$\psi\pi\rightarrow \psi\gamma$ (b)}, 
             \mbox{$\overline{\psi} \psi\rightarrow \sigma\gamma$ (c)} 
             and \mbox{$\psi\sigma\rightarrow \psi\gamma$ (d)}.
             Phase contour definitions as in Fig.~\ref{fig_Thermodyn}}
   \label{fig_rates}
\end{figure}

In \eqref{rate_allg}, four of the nine integrations can be carried out exactly applying the delta distribution. 
Another (angular) integration drops out by symmetry reasons, so one is left with four integrals, which have to be 
executed numerically resulting in the rates depicted in Fig.~\ref{fig_rates}.
When photon energies $\omega$ are much larger than the respective masses, it is not expected
to see much of the details of the phase structure. Contrary, at lower energies there are huge differences in available 
phase space and matrix elements squared leading to pronounced patterns which reflect phase diagram features, in particular
the effective masses. For this reason $\omega=10\MeV$ is chosen.

Figure~\ref{fig_rates} shows contour plots of the photon rates for the different contributing processes over the phase diagram.
In Figs.~\ref{subfig_rate_qp_gq_omega=0010} and \ref{subfig_rate_qs_gq_omega=0010} 
we see an enhancement in the 
crossover region and in Fig.~\ref{subfig_rate_qs_gq_omega=0010} a global maximum in the critical region.
In Figs.~\ref{subfig_rate_qq_gp_omega=0010} and ~\ref{subfig_rate_qq_gs_omega=0010} 
one notices
large rates in the chirally restored phase and much less photon emission in the chirally broken phase, which in case of
\ref{subfig_rate_qq_gs_omega=0010} is superimposed by an island of enhanced rates for $T\sim100\MeV$ and $\mu\lesssim200\MeV$.
Figures \ref{subfig_rate_qq_gp_omega=0010} and \ref{subfig_rate_qp_gq_omega=0010} show photon rates from processes involving
pions. Pions exhibit a large mass difference between the two 
phases, but contrary to the sigma meson whose mass has a global minimum at the CP
the pion mass does not show special features at this point. This leads to 
a large difference in the emissivity between the phases but no features characteristic for the CP itself.
For the pion-involving Compton process (Fig.~\ref{subfig_rate_qp_gq_omega=0010}) there is an enhancement in the 
crossover region. This is probably due to a combination of phase space
effects and the (comparatively) large probability for the internally propagating pion to get on-shell.
A better channel for obtaining signatures of a CP are sigma involving processes. This is expected, since the sigma meson
is precisely the mode getting massless at the CP making long range interactions possible
and thus driving the critical processes. Unfortunately, the inclusion
of linearized fluctuations increases the sigma mass, so it is not clear whether the endpoint of the 1st order phase transition
shows correctly the critical behavior. But linearizing fluctuations anyhow restricts to small fluctuations making it not adequate
very near the CP. Nevertheless the sigma mass drops to small values in the critical
region, which has a notable effect on the corresponding processes, \eg the excess of the photon rate in the critical region 
in Fig.~\ref{subfig_rate_qs_gq_omega=0010}. 

There is a large difference in the rates for the processes
under consideration, even between the corresponding Compton (Figs.~\ref{fig_rates}(b) and \ref{fig_rates}(d)) and annihilation 
(Figs.~\ref{fig_rates}(a) and ~\ref{fig_rates}(c)) processes. These can be understood in terms
of available phase space in combination with thermal suppression. Within a Boltzmann approximation two of the remaining integrals
in \eqref{rate_allg} can be solved to obtain
\begin{eqnarray}
   \omega \frac{d^7 N}{dx^4dk^3} 
        &\stackrel{\omega\ll m_i}{\sim}&\int \limits_{s_0}\frac{ds}{s-z^2}\int dt |M(s,t)|^2 \exp\{-(s-z^2)/(4\omega T)\}.
\end{eqnarray}
The difference between the minimal kinematically allowed value of the center of mass energy 
$\sqrt{s_0} = \max\{m_1+m_2, m_3\}$ 
for the different processes, together with a small value of $\omega$
leads to the huge thermal suppression at small $T$ seen in Figs.~\ref{fig_rates} (a) and (d).
\section{Summary}
Focusing on soft-photon emission rates we demonstrate that some features of the phase diagram provided by the 
linear sigma model are nicely mapped out. Being aware of some limitations, such as the restriction to linearized 
fluctuations (\cf \cite{Tripolt:2013jra} for a proper account of fluctuations) and the need to implement more complete 
rates in a model of space-time evolution of the matter, we hope that improved calculations can provide useful complementary
information on strongly interacting matter produced in the course of relativistic heavy-ion collisions at various energies, 
system sizes and centralities.

\end{document}